\documentclass[preprint,onecolumn,nofootinbib]{revtex4}
\usepackage[colorlinks=true,linkcolor=blue,urlcolor=blue,filecolor=black,citecolor=red,pdfstartview=FitV,pdftitle={},pdfsubject={},pdfkeywords={},pdfpagemode=None,bookmarksopen=true]{hyperref}
\usepackage{graphicx}
\usepackage{amsmath}
\usepackage{amssymb,ulem}
\usepackage{color}%
\usepackage{tikz}
\usepackage{dcolumn}
\usepackage{bbold}
\usepackage{times}
\usepackage{xcolor}
\usepackage{setspace}

\begin{document}
\title{Dynamical descalarization in Einstein-Maxwell-scalar theory}
\author{Chao Niu $^{1}$}
\email{niuchaophy@gmail.com}
\author{Wei Xiong $^{1,2}$}
\email{phyxw@stu2019.jnu.edu.cn}
\author{Peng Liu $^{1}$}
\email{phylp@email.jnu.edu.cn}
\author{Cheng-Yong Zhang $^{1}$}
\email{zhangcy@email.jnu.edu.cn}
\author{Bin Wang $^{3,4}$}
\email{wang\_b@sjtu.edu.cn}
\address{\textit{1. Department of Physics and Siyuan Laboratory, Jinan University, Guangzhou 510632, China}} 
\address{\textit{2. School of Physics and Optoelectronics, South China University of Technology, Guangzhou 510641, People’s Republic of China}}
\address{\textit{3. Center for Gravitation and Cosmology, College of Physical
      Science and Technology, Yangzhou University, Yangzhou 225009, China}}
\address{\textit{4. School of Aeronautics and Astronautics, Shanghai Jiao Tong
      University, Shanghai 200240, China}}

\begin{abstract}

    
For an asymptotically flat hairy black hole in the Einstein-Maxwell-scalar (EMS) theory, we study the possibility of shedding off its scalar hair via nonlinear scalar perturbation fully interacting with the background spacetime. We examine the  effect of the perturbation strength on the descalarization. The results show that the effective charge to mass ratio of the black hole plays the key role in the dynamical descalarization. The descalarization at the threshold is continuous. This indicates a second order phase transition.
\end{abstract}
\maketitle

\section{Introduction}

In the framework of general relativity (GR), we have the black hole no hair theorem which states that a  black hole can be   described by only three macroscopic degrees of freedom (mass $M$, electric charge $Q$ and angular momentum $J$).  In order to overcome problems in GR, such as the non-renormalization  and the inevitable singularities etc., modified theories have been introduced to study the gravity.  In the modified gravity, it was found that hairy black holes with more degrees of freedom can exist  \cite{Herdeiro2015,Volkov2015}, for example in the Einstein-dilaton-Gauss-Bonnet theory \cite{Torii1997,Kanti1997,Kleihaus2011,Kleihaus2015,Pani2011,Herdeiro2014,Ayzenberg2014}, or with other matter fields \cite{HHVolkov1989,HHBizon1990,HHGreene1993,HHMaeda1994,HHLuckock1986,HHDroz1991}.  
In this paper, we focus on the  black hole solutions with non-trivial scalar hair. They can be generated by the self-interaction potential of the scalar field \cite[]{Herdeiro2015,Zou:2014sja,Mai:2020sac,Hong:2020miv,Mayo:1996mv,Collodel:2022jly}. For a Kerr black hole, the mass term of the scalar field induces  superradiant instability and creates a stationary scalar hair around the black hole \cite[]{Cardoso:2005vk,Dolan:2007mj,Hod:2012px,Herdeiro:2014goa}. The dilatonic coupling between the scalar field and other source terms (curvature term or electromagnetic field) can also permit the hairy black hole solutions \cite[]{Torii1997,Kanti1997,Kleihaus2011,Kleihaus2015,Pani2011,Herdeiro2014,Ayzenberg2014,Zhang2021j,Zhang2021a,Richarte2022,Zhang:2015jda}. 

There exists a dynamical process, called spontaneous scalarization\cite{Damour1993,Damour1996,Harada1997,Cardoso2013,Cardoso2013b,Zhang2014}, to address scalar hair onto bald black holes and transform them into hairy black holes \cite{Doneva1711,Herdeiro2018,Fernandes2019,Silva1711,Antoniou1711,Cunha1904,Dima:2020yac,Herdeiro2009,Berti2009,Brihaye2018,Guo2021,Ripley2020,Ripley2021,Ripley2020b,Kuan2021},  branching from the Reissner-Nordstr\"om (RN) or Schwarzschild black holes.
In this paper we focus on the Einstein-Maxwell-scalar (EMS) theory which induces the spontaneous scalarization through a non-minimal coupling between the scalar field and the electromagnetic field \cite[]{Herdeiro2018,Fernandes2019,Guo2021}. This model allows the electro-vacuum solution (the RN black hole), in different from the  Einstein-Maxwell-dilaton  theory \cite[]{Zhang2021j,Zhang2021a,Richarte2022}. Under an arbitrary small perturbation, the RN solution in this model spontaneously evolves to the hairy solution and the corresponding dynamic evolution was investigated in \cite{Xiong:2022ozw,Zhang2021b,Luo:2022roz} for different asymptotic behaviors of spacetimes. In \cite{Xiong:2022ozw} we confirmed that the growth rate of scalar hair for spontaneous scalarization is consistent with the fundamental instability mode calculated in the linear perturbation theory. 

In addition to the scalarization, scalar hair on hairy black holes can also be removed through descalarization processes. The dynamic process of black hole descalarization was  studied in \cite{Silva2020,Doneva:2022byd,Elley:2022ept} for the scalar-Gauss-Bonnet (sGB) theory where the removal of scalar hair is realized by the head-on collision of binary black holes. However, in these articles, the backreaction of scalar to the spacetime is not taken into account, where the nonlinear scalar equation of motion was only examined on the background.  Another descalarization mechanism induced by accretion was also studied in  generalized scalar-tensor theory  \cite{Kuan2022,Liu:2022eri,Liu:2022fxy}.  In EMS theory, the dynamical  descalarization was studied in asymptotically anti-de Sitter (AdS) spacetime \cite{Zhang:2022cmu}. It was found that the descalarization can be either continuous in models allowing spontaneous scalarization  or
discontinuous in models allowing nonlinear scalarization at the threshold. The descalarization has  attracted more and more attention recently due to its potential observational signature in the gravitational wave signal \cite{Silva2020, Kuan2022,Doneva:2022byd,Zhang:2022cmu,Liu:2022eri,Elley:2022ept,Liu:2022fxy}. 

To simulate the astrophysical environment, we concentrate our attention in an asymptotically flat spacetime and consider the fully nonlinear dynamical descalarization process of shedding off scalar hair from  a spherically symmetric scalarized black hole in the EMS theory. 
In EMS theory, hairy black holes exist only when charge to mass ratio is high and the scalar field is strongly coupling to the Maxwell field.  
If the final black hole mass increases so that the charge to mass ratio drops, the scalar hair can be deprived.  In this work we will exhibit that such descalarization process can happen for a single black hole absorbing enough energy.

The organization of the paper is as follows. In section \ref{section II} we will introduce the EMS model, write out the equations of motion and discuss the initial conditions. Then in \ref{section III} we will present numerical result on the  descalarization and disclose the relation between the energy dissipation and descalarization. Finally in the last section, we will summarize the obtained results.

\section{The EMS model}
\label{section II}

The action of the Einstein-Maxwell-scalar theory 
\begin{equation}
    S=\frac{1}{16\pi}\int d^{4}x\sqrt{-g}\left[R-2\nabla_{\mu}\phi\nabla^{\mu}\phi-f(\phi)F_{\mu\nu}F^{\mu\nu}\right],
    \label{eq:action}
\end{equation}
where $R$ is the Ricci scalar, $\phi$ is the scalar field, $F_{\mu\nu}=\nabla_{\mu}A_{\nu}-\nabla_{\nu}A_{\mu}$ and $A_{\mu}$ is the Maxwell gauge field.
In this work, we take an exponential coupling
\begin{equation}
    f(\phi) = e^{-b\phi^{2}},\ \  b <0,
\end{equation}
which remains a $\mathbb{Z}_{2}$ symmetry $\phi \rightarrow -\phi$ in the action (\ref{eq:action}). The equations of motion for the gravity, scalar and Maxwell field are respectively 
\begin{align}
    R_{\mu\nu}-\frac{1}{2}Rg_{\mu\nu}= & 2\left[\partial_{\mu}\phi\partial_{\nu}\phi-\frac{1}{2}g_{\mu\nu}\nabla_{\rho}\phi\nabla^{\rho}\phi+f(\phi)\left(F_{\mu\rho}F_{\nu}^{\ \rho}-\frac{1}{4}g_{\mu\nu}F_{\rho\sigma}F^{\rho\sigma}\right)\right],\\
    \nabla_{\mu}\nabla^{\mu}\phi= &-\frac{b}{2} \: e^{-b\phi^{2}}F_{\mu\nu}F^{\mu\nu} \phi \label{eq:equations of scalar},\\
    \nabla_{\mu}\left(f(\phi)F^{\mu\nu}\right)= & 0.
    \label{eq:equations of motion}
\end{align}
 The non-minimal coupling $f(\phi)$ between the scalar field and the Maxwell field   provides the scalar field a negative effective mass, which triggers the  tachyonic instability of the scalar field perturbation in the RN black hole background in some parameter region.

\subsection{The nonlinear equations}
\label{subsection II A}
To simulate the nonlinear  descalarization of shedding of scalar hair from  a  spherically symmetric hairy black hole in the EMS model, we use the Painlev\'{e}-Gullstrand (PG)-like coordinate ansatz
\begin{equation}
    ds^{2} =-\left(1-\zeta^{2}\right)\alpha^{2}dt^{2}+2\zeta\alpha dtdr+dr^{2}+r^{2}(d\theta^{2}+\sin^{2}\theta d\phi^{2}).
    \label{eq:metric ansatz}
\end{equation}
This coordinate is regular on the horizon and allows one of the radial boundaries of the numerical evolution to be within the black hole horizon. We take the   gauge field as
\begin{equation}
    A_{\mu}dx^{\mu}=A(t,r)dt,
\end{equation}
then the Maxwell equation (\ref{eq:equations of motion}) gives $\frac{1}{\alpha}\partial_{r}A=\frac{Q}{r^{2}f(\phi)}$, in which $Q$ is interpreted as the electric charge. For numerical simulation 
we introduce auxiliary variables
\begin{equation}
    \Phi=\partial_{r}\phi,\ \ \ P=\frac{1}{\alpha}\partial_{t}\phi-\zeta\Phi.
    \label{eq:auxiliary variables}
\end{equation}
The equations of motion become
\begin{align}
    \partial_r\alpha= & -\frac{rP\Phi\alpha}{\zeta}, \label{eq:alpha}\\
    \partial_r\zeta= & \frac{r}{2\zeta}\left(\Phi^{2}+P^{2}+\Lambda\right)+\frac{Q^{2}}{2r^{3}\zeta f(\phi)}+rP\Phi-\frac{\zeta}{2r},\label{eq:zetadr}\\
    \partial_{t}\zeta= &\frac{r\alpha}{\zeta}\left(P+\Phi\zeta\right)\left(P\zeta+\Phi\right).\\
    \partial_{t}\phi =& \alpha\left(P+\Phi\zeta\right),\label{eq:phit}\\
    \partial_{t}P   =&\frac{\left(\left(P\zeta+\Phi\right)\alpha r^{2}\right)'}{r^{2}}+\frac{\alpha}{2}\frac{f'(\phi)Q^{2}}{r^{4}f^{2}(\phi)} 
    \label{eq:equations of zeta and phi}
\end{align}
$\alpha$ and $\zeta$ can be solved directly by a given $\phi,P$ from equation (\ref{eq:auxiliary variables}, \ref{eq:alpha}) and   (\ref{eq:zetadr}). We use them to obtain the initial profile and the detailed process will be explained in     subsection \ref{subsection 2.B}.

The boundary conditions for the nonlinear evolution are given by
\begin{equation}
    \alpha|_{r \rightarrow \infty} =1 \ , \ \ \zeta|_{r \rightarrow \infty} = \sqrt{\frac{2M}{r}} \ , \ \ \phi|_{r \rightarrow \infty}=0.
    \label{eq:boundary condition}
\end{equation}
Here $M$ is the Arnowitt-Deser-Misner (ADM) mass, which includes the energy of the gravity, Maxwell and scalar fields in the system \cite{Hayward:1994bu}. The first condition implies that we take the time coordinate $t$ as the proper time of the observer at spacial infinity. We will use the Misner-Sharp mass defined as
\begin{equation}
M_{MS}(t,r)=\frac{r}{2}\left(1-g^{\mu\nu}\partial_\mu{\zeta}\partial_\nu{\zeta}\right)=\frac{r}{2}\zeta(t,r)^{2},
\label{eq:Misner Sharp mass}
\end{equation}
which tends to the total mass $M$ when $r \rightarrow \infty$. We also define the  mass density 
\begin{equation}
    m_{MS}(r)= \frac{d\: M_{MS}(r)}{dr}
    \label{eq:Misner Sharp mass density}
\end{equation}
to exhibit the mass distribution of spacetime to better reveal the differences between the scalarized black hole and the scalar-free black hole. The reduced horizon area is given by
\begin{equation}
    a_{h}=\frac{A_{h}}{16 \pi M^{2}}
\end{equation}
where the $A_{h}$ is the area of the apparent horizon. 

\subsection{Spontaneous scalarization}
\begin{figure}[htbp]
    \centering
    \includegraphics[width = 0.45\textwidth]{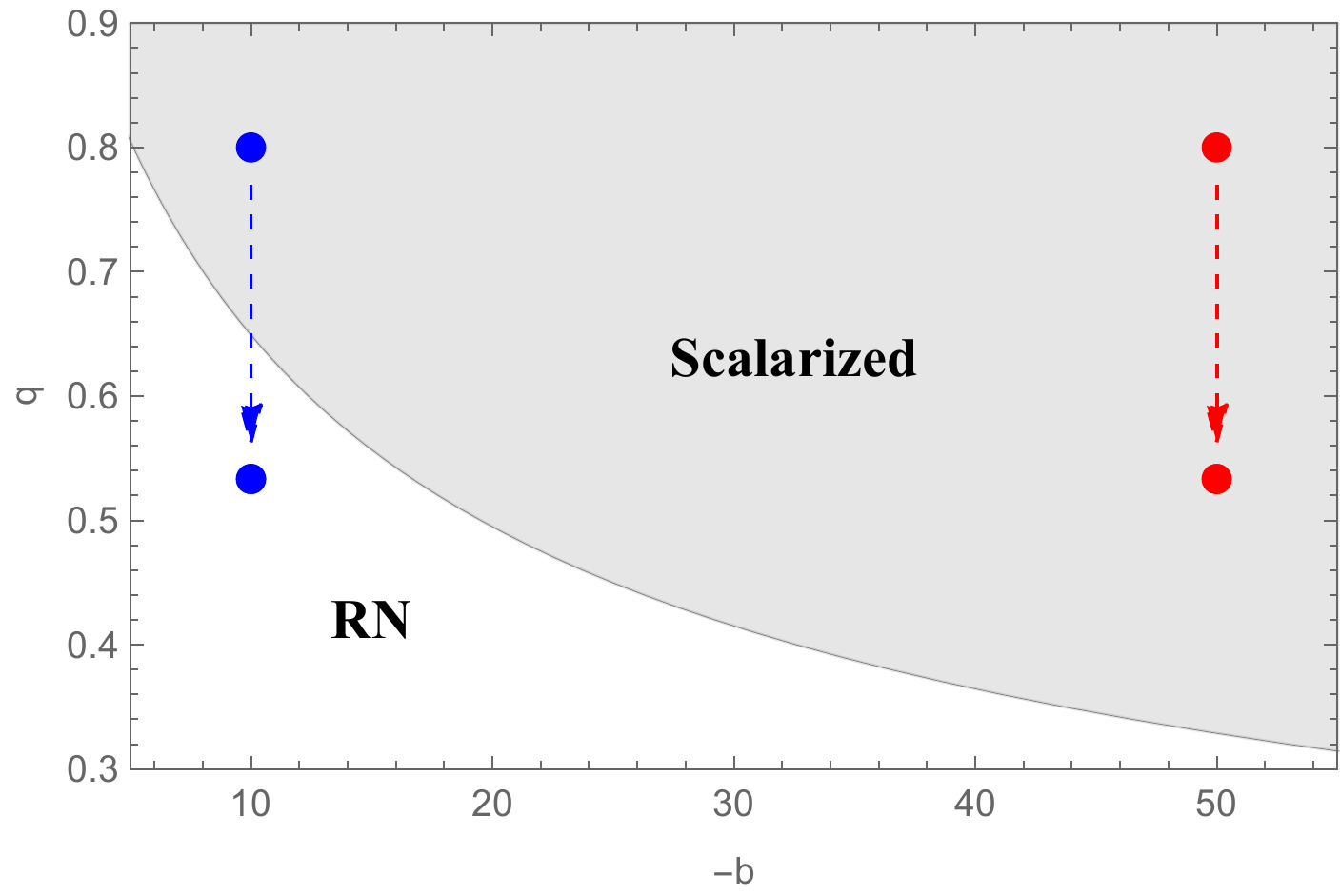}
    \caption{The space of parameters for the existence of bald and scalarized black holes. $q$ is  the charge to mass ratio $Q/M$ and $b$ is the coupling strength. The scalarized black hole is stable in the gray region. For small fixed value of $b$, it is possible to transform a hairy black hole to a bald black hole through the decrease of $q$, as indicated in the blue arrow.  }
    \label{fig:migration}
\end{figure}

\begin{figure}[htbp]
    \centering
    \includegraphics[width = 0.4\textwidth]{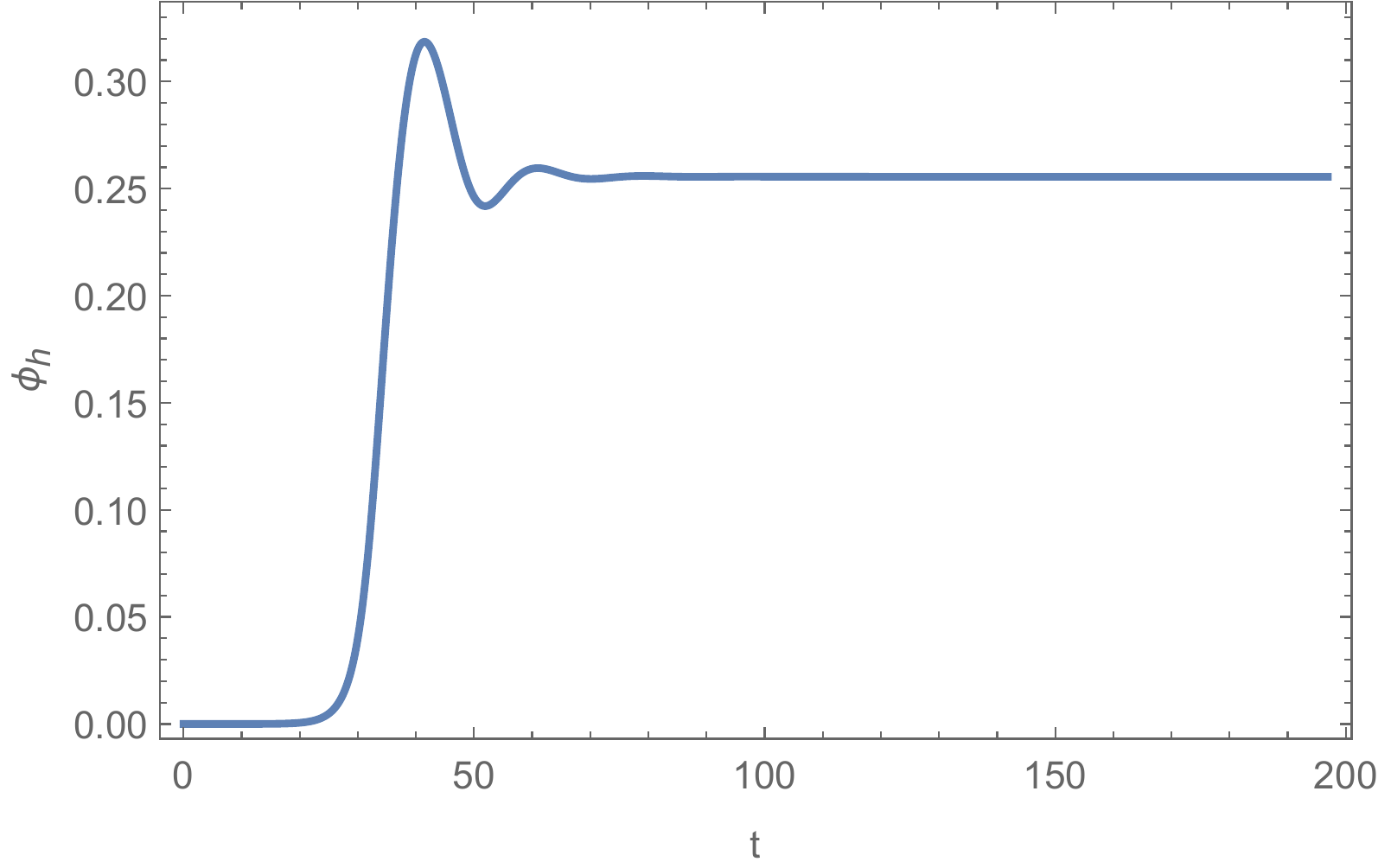}
    \includegraphics[width = 0.4\textwidth]{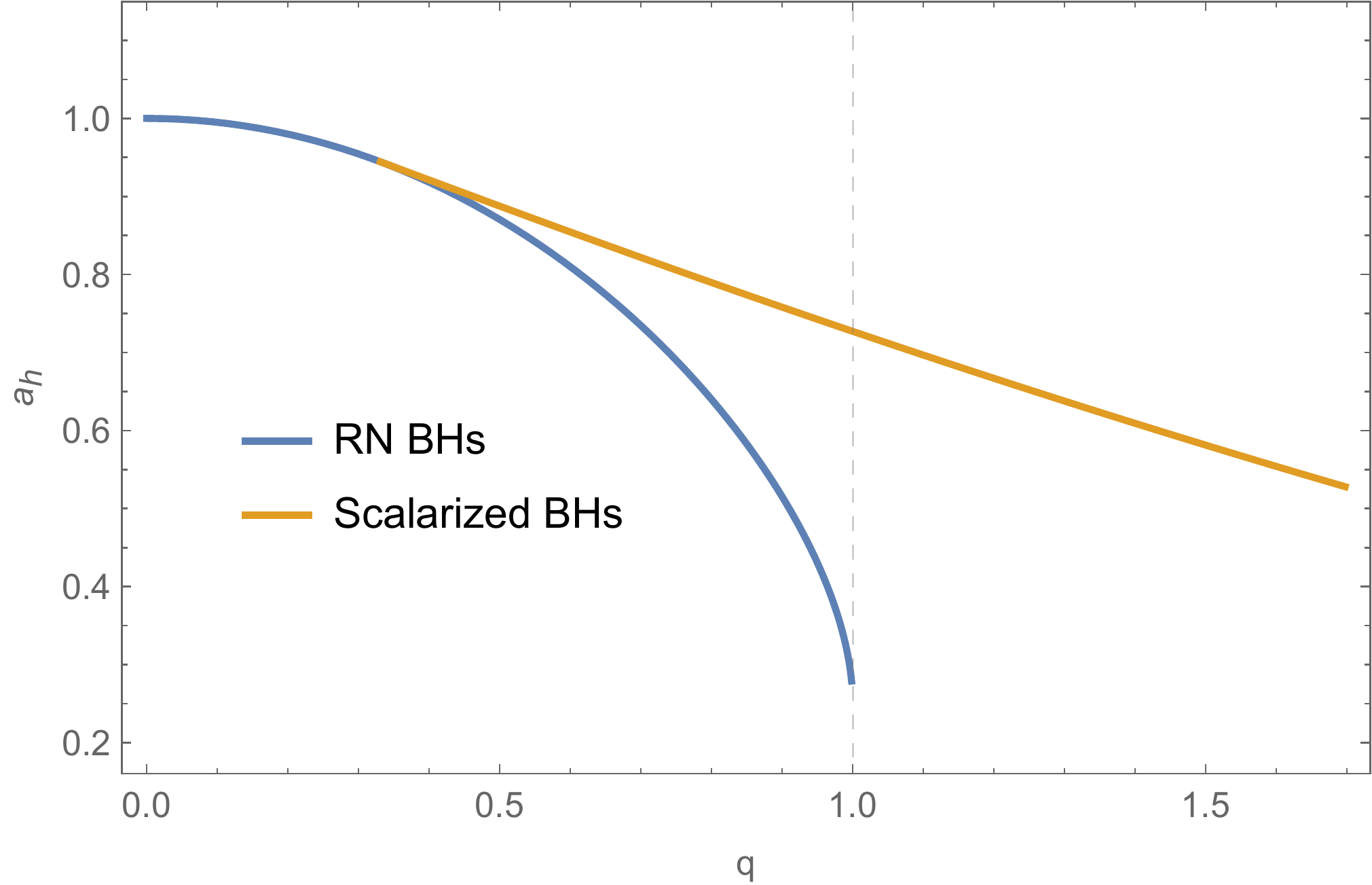}
    \caption{Left: the dynamical spontaneous scalarization of the RN black hole with $M=1,Q=0.8,-b=50$ under a tiny perturbation. Right: the behaviors of reduced horizon areas $a_{h}$ for a static scalarized black hole and the RN black hole when $-b=50$. }
    \label{fig:scalarization example}
\end{figure}

We first solve the equations of motion and display the space of parameters for the existence of scalarized black holes in the EMS theory in Fig.\ref{fig:migration}. The results are consistent with \cite{Herdeiro2018}. In the gray region,  the tachyonic instability of the scalar field in the bald RN black hole background is triggered. At the nonlinear level, while the  RN black hole  is perturbed by an arbitrarily small scalar perturbation, the scalar field exponentially grows and converges to a certain value, as shown in the left panel of Fig.\ref{fig:scalarization example}. In the final state, the black hole is covered by static scalar hair. 

The right panel of Fig.\ref{fig:scalarization example} shows that the reduced horizon area of the scalarized black hole branches off the RN black hole, and stays bigger. From the perspective of black hole thermodynamics, this tells us that a hairy black hole  contains larger entropy than a bald black hole. At a threshold value of the  charge to mass ratio,  the thermodynamic instability can destroy the scalar-free black hole and transform it to a hairy black hole through a phase transition. For the hairy black hole, its charge to mass ratio can exceed the maximum value of the RN hole.

\subsection{Initial conditions for descalarization}

\label{subsection 2.B}

\begin{figure}[htbp]
    \centering
    \includegraphics[width = 0.4\textwidth]{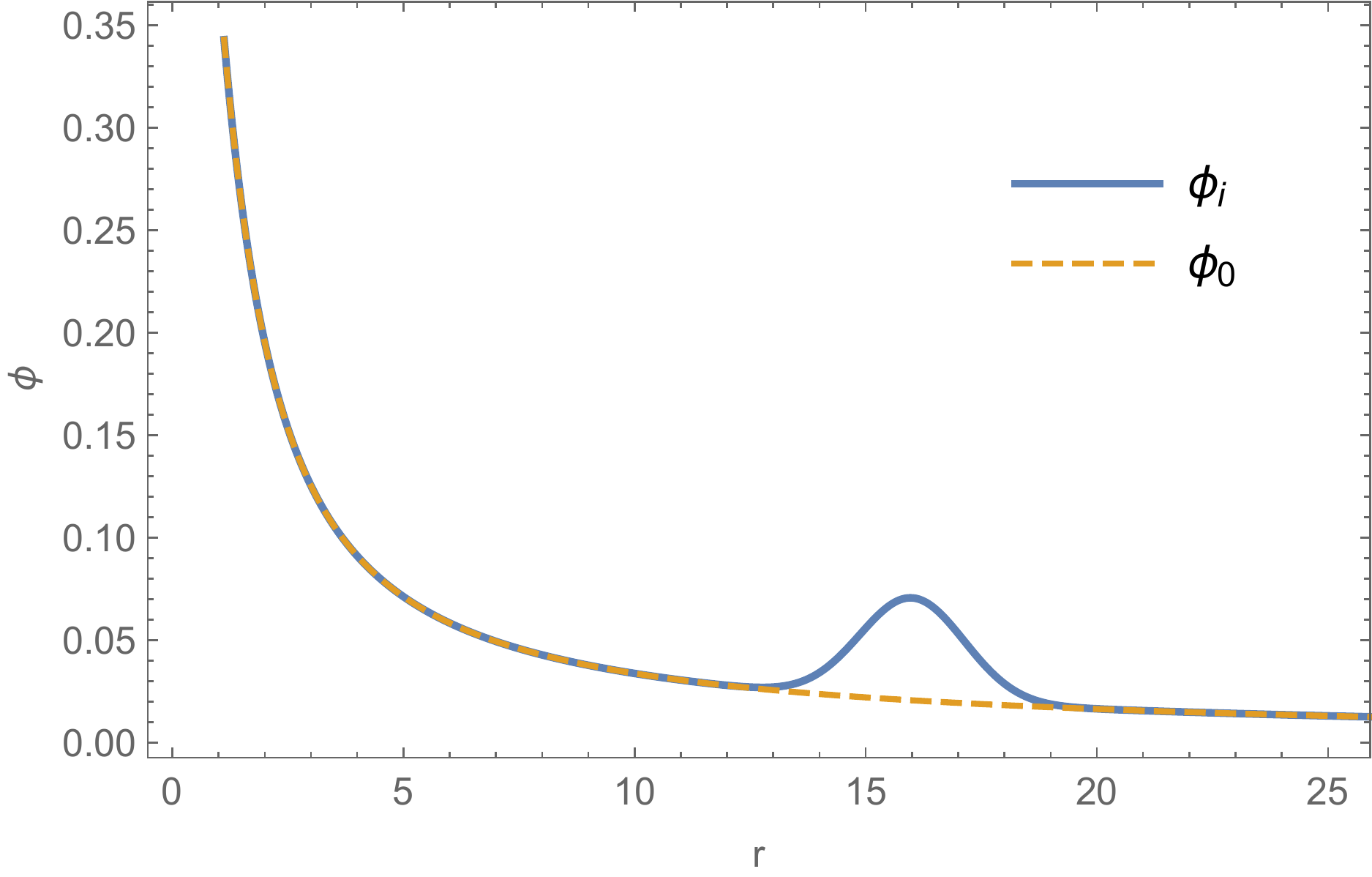}
    \includegraphics[width = 0.4\textwidth]{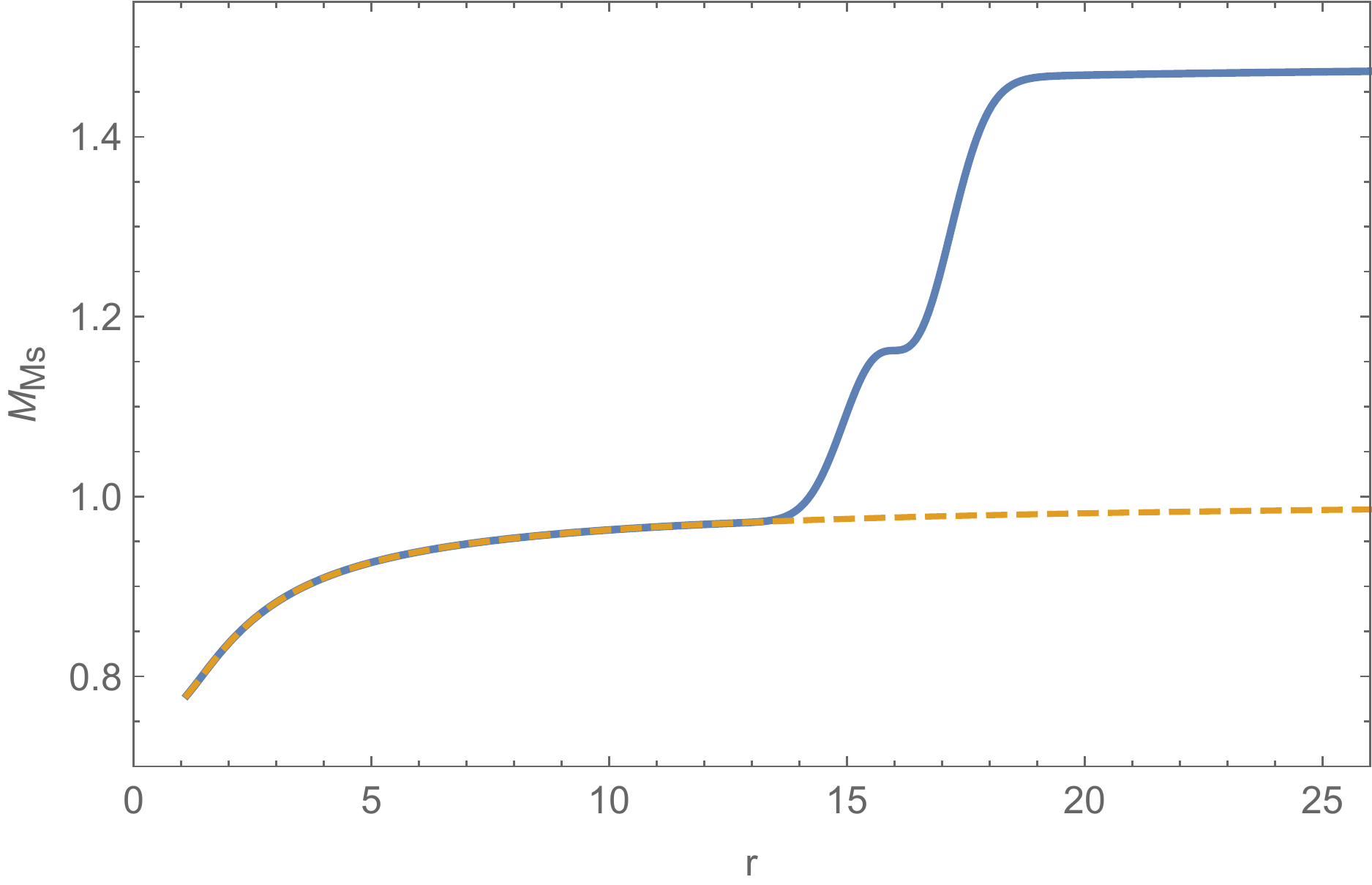}
    \caption{Left: the initial data $\phi_{i}$ on the background $\phi_{0}$  of the initial scalarized black hole. Right: the distribution of the Misner-Sharp mass for the background scalarized black hole (yellow) and the initial black hole solution after a finite perturbation (blue).}
    \label{fig:initial profile}
\end{figure}

In this subsection, we discuss the initial condition for studying the dynamical descalarization. 
 We add a wave packet of  scalar field perturbation to the initial scalarized black hole, as shown in the left panel of Fig.\ref{fig:initial profile}. Specifically,  the initial scalar pulse is  given by
\begin{align}
    \phi_{i} = & \phi_{0} + 10^{-2}B\: e ^{-\left( \frac{r-10 \: r_{h}}{r_{h}}\right)^{2}}, \label{eq:phii} \\
    P_{i}    = & \pm  \frac{20 r_{h}-2 r}{r_{h}^{2}}  10^{-2}B\: e ^{-\left( \frac{r-10 \: r_{h}}{r_{h}}\right)^{2}} - \zeta_0\Phi_0, \label{eq:Pi}
\end{align}
in which $\phi_{0},\zeta_0,r_h$ are the  solutions of the scalar field, metric function and the  horizon radius  of the background scalarized black hole, respectively.  $B$ is the strength of the initial perturbation and $\Phi_0=\partial_r \phi_0$. 
The sign $\pm$ in (\ref{eq:Pi})  implies an ingoing/outgoing initial pulse respectively.  

Given an initial   perturbation, one can work out the initial metric functions $\zeta,\alpha$ by solving constraint equations (\ref{eq:alpha},\ref{eq:zetadr}) with an appropriate boundary condition  at the spatial infinity. According to the  relation between $\zeta$ and the Misner-Sharp mass (\ref{eq:Misner Sharp mass}), we can describe this operation from the perspective of the Misner-Sharp mass. From   (\ref{eq:boundary condition}) the parameter $M$, which is the asymptotic value of Misner-Sharp mass at infinity, can be divided into two part
\begin{equation}
    M = M_0 + \Delta M. \label{eq:MdM}
\end{equation}
The $M_0$ is the mass of the original scalarized black hole system which we fixed as $M_0=1$. The $\Delta M$ is the additional mass from the wave packet. The energy $\Delta M$ is positively related to the  strength of the wave packet. 

The right panel of Fig.\ref{fig:initial profile} sketches the behavior of the Misner-Sharp mass for  appropriate parameters $\Delta  M, B$. Although initially the blue line completely coincides with the yellow line for the original $\zeta$ of the scalarized black hole, at the position of the wave packet we observe a rapid jump in the blue line because of the addition of energy from the scalar perturbation.  

The evolution is examined by using the fourth order difference method in the radial direction and the fourth order Runge-Kutta method in the time direction. We work out in the radial region from $r_{0}$ to $\infty$, where $r_{0} = 1.48$ always lies in the apparent horizon during the evolution. The  radial space is compatified in a  region  $(z_{0},1)$ by a coordinate transformation $z = r/(r+1)$. More details were discussed in \cite{Xiong:2022ozw}. 

\section{Numerical results}
\label{section III}

\subsection{Descalarization}

\begin{figure}[htbp]
    \centering
    \includegraphics[width = 0.4\textwidth]{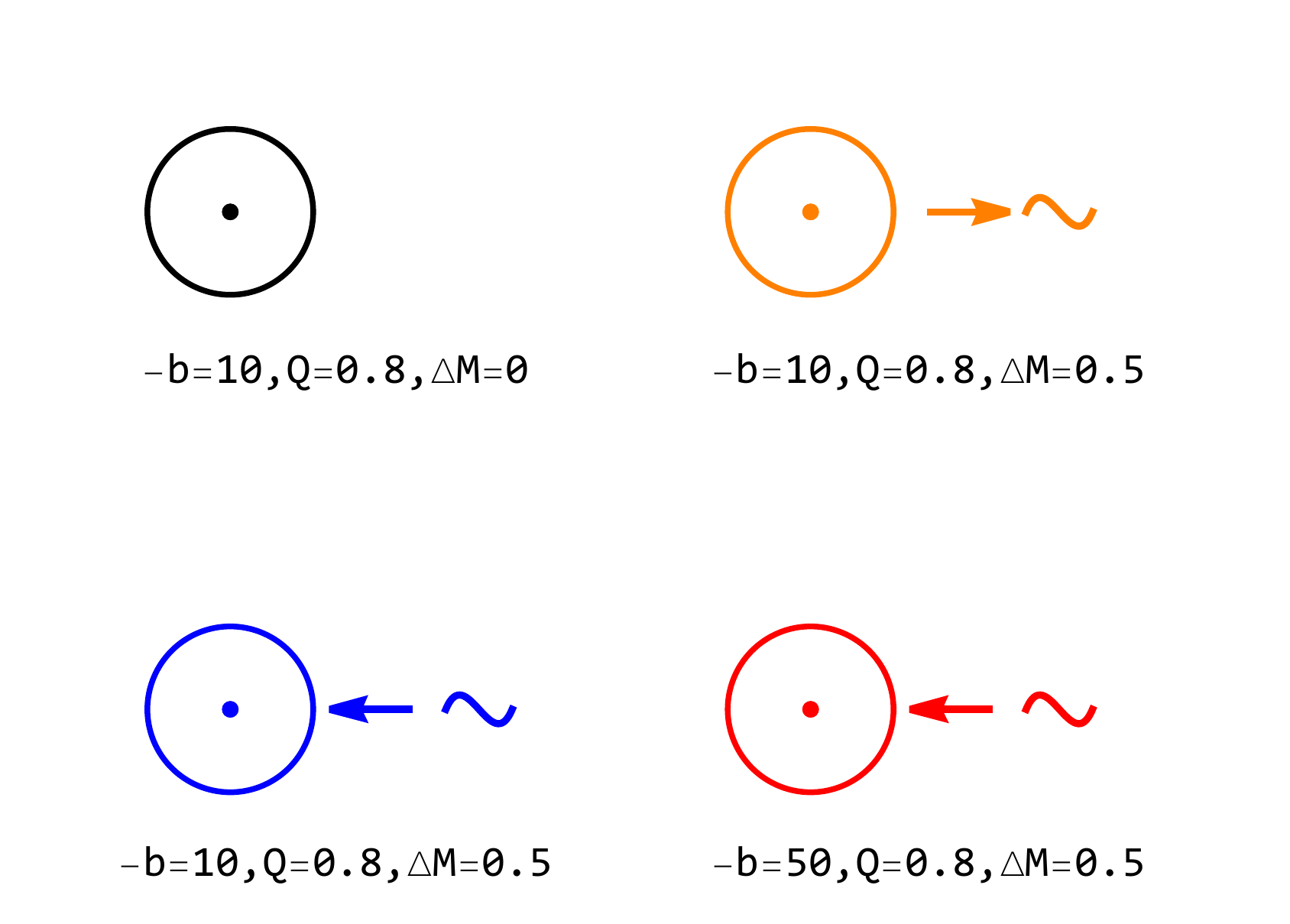}
    \includegraphics[width = 0.4\textwidth]{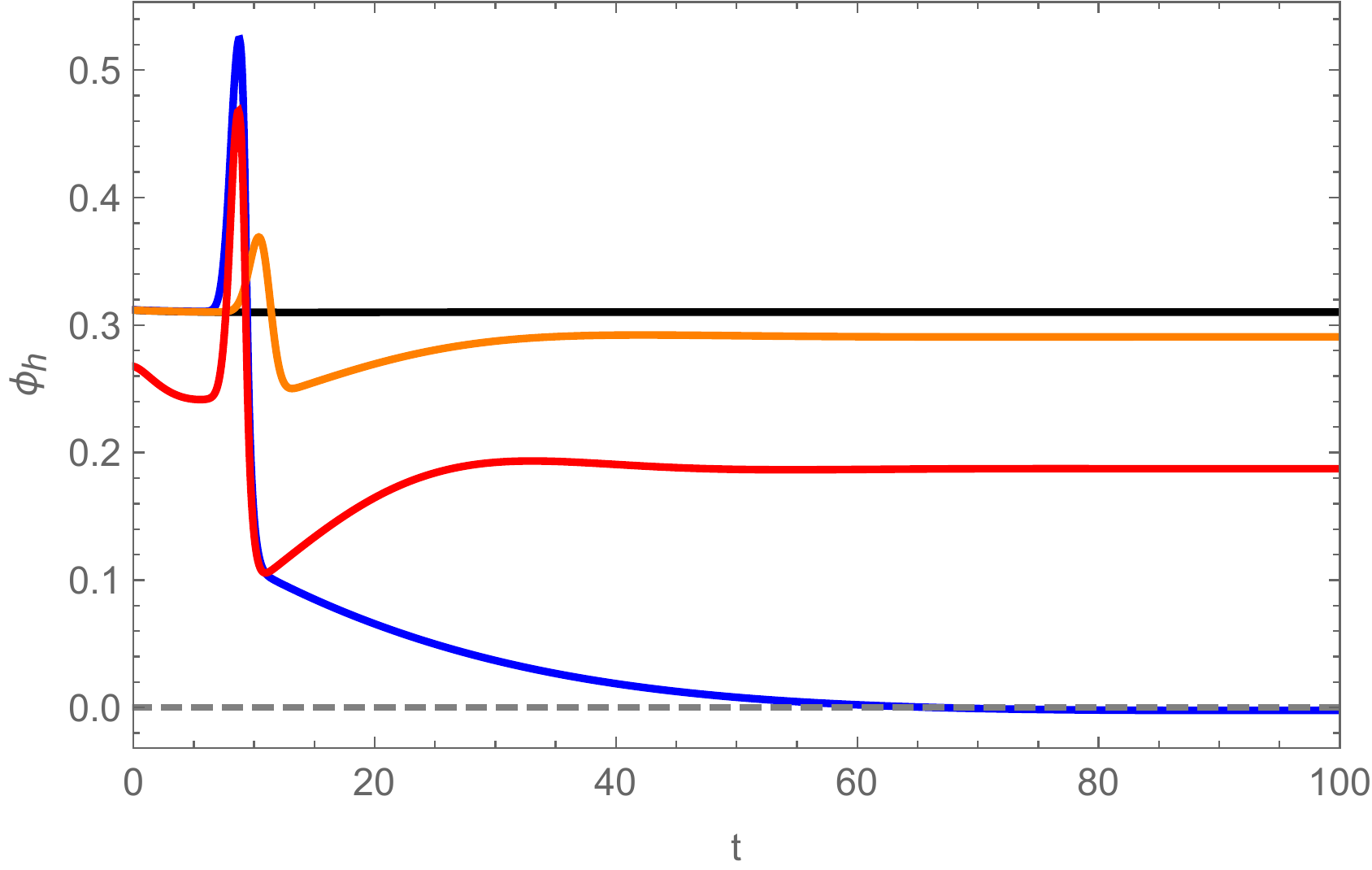}
    \includegraphics[width = 0.4\textwidth]{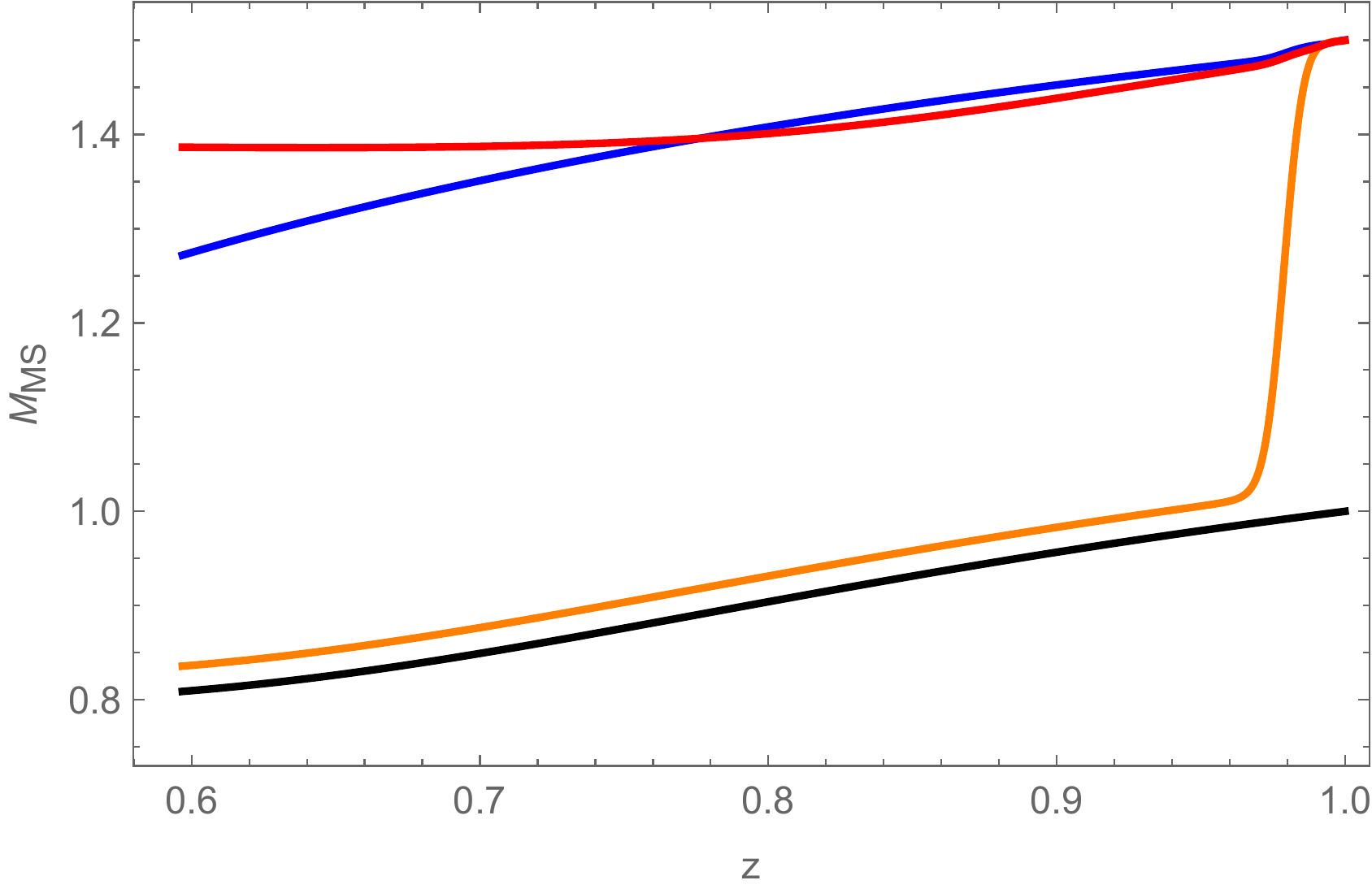}
    \includegraphics[width = 0.4\textwidth]{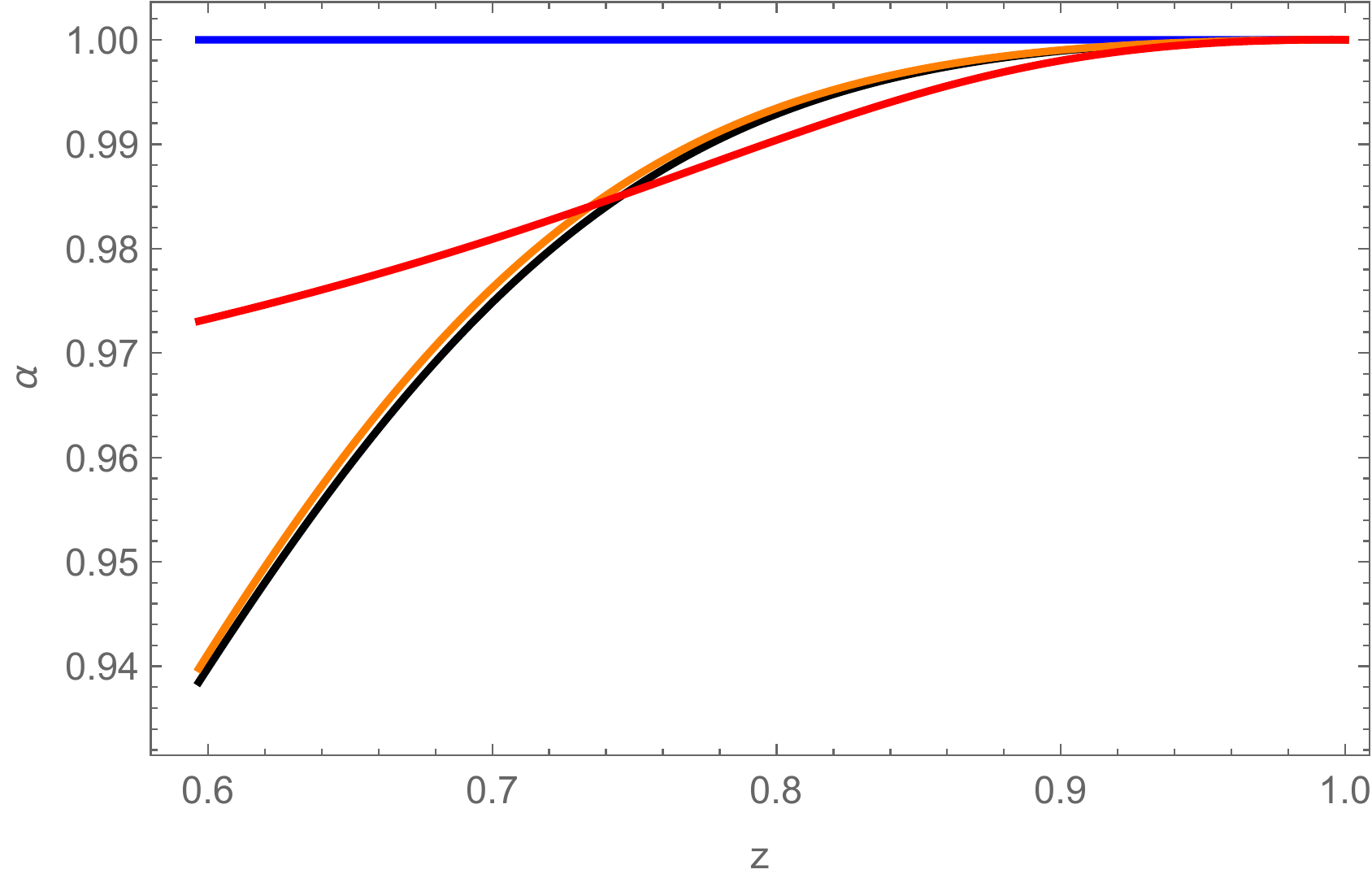}
    \caption{Four types of evolution are displayed in different colors and their parameters are listed in the upper left panel. The remaining panels include the time evolution for the scalar field at the horizon (upper right), and the Misner-Sharp mass distribution (bottom left), the profile of $\alpha$ (bottom right) at late times, respectively.}
    \label{fig:category}
\end{figure}

In this subsection, we discuss the process of the descalarization. We first confirm the validity of our program. As shown  in Fig.\ref{fig:category},  a scalarized black hole without scalar perturbation keeps unchanged. 
The orange line describes a scalarized black hole with an outgoing scalar perturbation and the blue line is for a scalarized black hole with an infalling of scalar wave packet respectively. 

The blue line in the upper right panel manifests that the scalar hair is deprived by the ingoing wave packet after a certain period of time.  Most energy of the wave packet is absorbed by the scalarized black hole and then $q$ decreases below the critical value $q_{\textrm{c}}$ as indicated in Fig.\ref{fig:migration}. This leads the hairy black hole finally evolves into a bald RN black hole.

The situation for the outgoing wave packet is the opposite. Only very small amount of energy is absorbed by the black hole.  Almost all energy of the wave packet escapes to the infinity and thus a cliff-like elevation appears in the orange line  for the  Misner-Sharp mass at the far region ($z\approx 1$). The escaping energy has little impact on the black hole system, which is clear by comparing the black and orange lines. As a result, the scalar hair is only a little smaller than the original scalarized black hole.

The results of choosing different coupling strength $-b=50$ are shown by the red line in Fig.4.  $q$ is still above $q_{\textrm{c}}$ when $-b=50$, so that the hairy black hole survives but with less scalar hair. The critical value $q_{\textrm{c}}$  decreases with the increase of $|b|$ as shown in Fig.1.  
 
The diverse evolution results with three initial conditions shown in blue, orange, red lines have the same total mass $M$. 
The asymptotic expression for the Misner-Sharp mass and $\alpha$ at far region is given by $M_{\textrm{s}}-Q^{2}/2r$ and $1$, respectively.  
We use interpolation to fit the first $80\%$ of the data of the Misner-Sharp mass to obtain the effective mass $M_{\textrm{s}}$ of each black hole system. All systems we simulated here has the same total mass $M=1.5$. But the values of $M_{\textrm{s}}$ are 1.000 (black), 1.025 (orange), 1.488 (blue), 1.486 (red) respectively. The remaining energy are carried by the scalar field escaping to the infinity.

\begin{figure}[htbp]
    \centering
    \includegraphics[width = 0.4\textwidth]{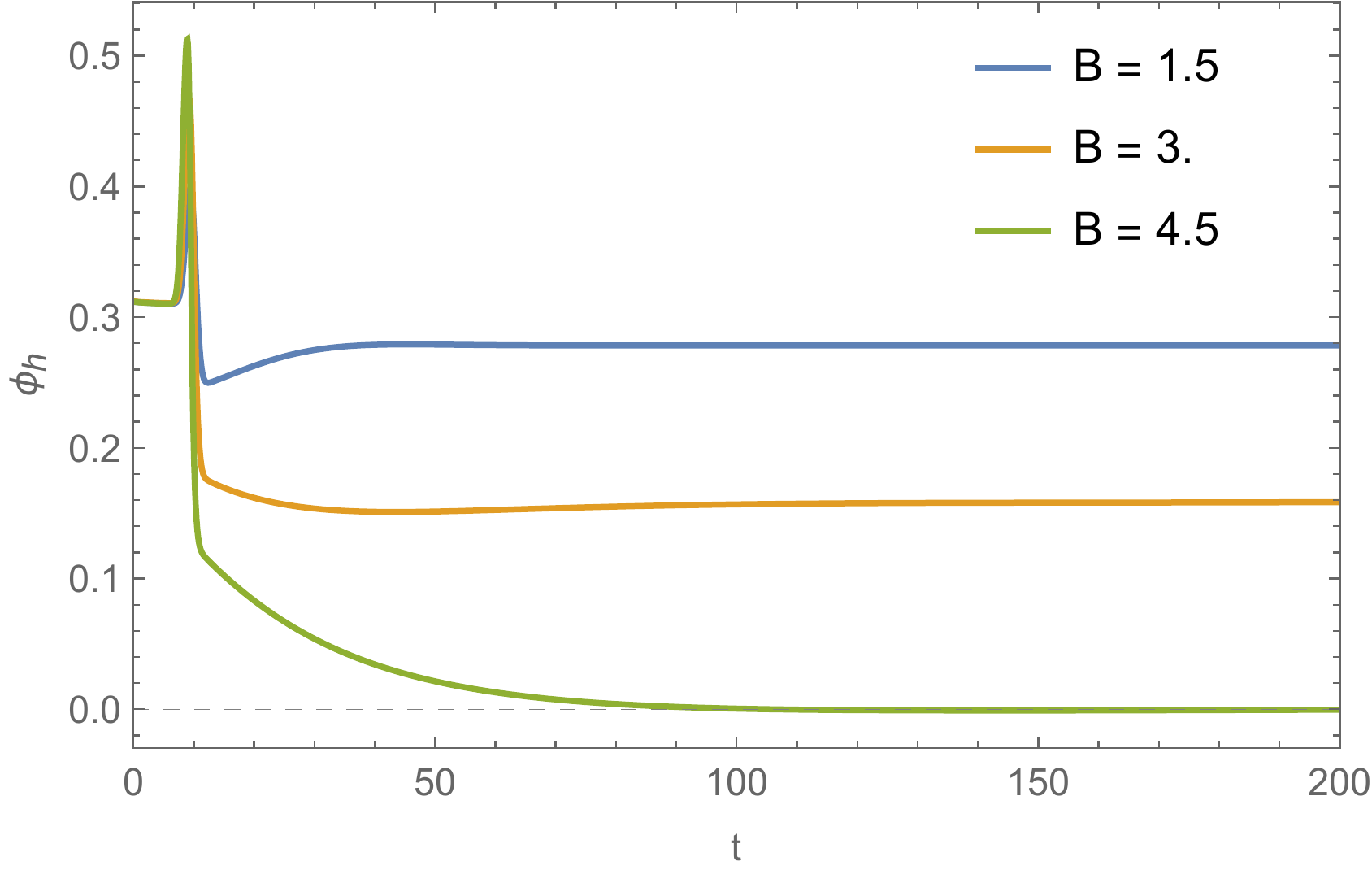}
    \includegraphics[width = 0.4\textwidth]{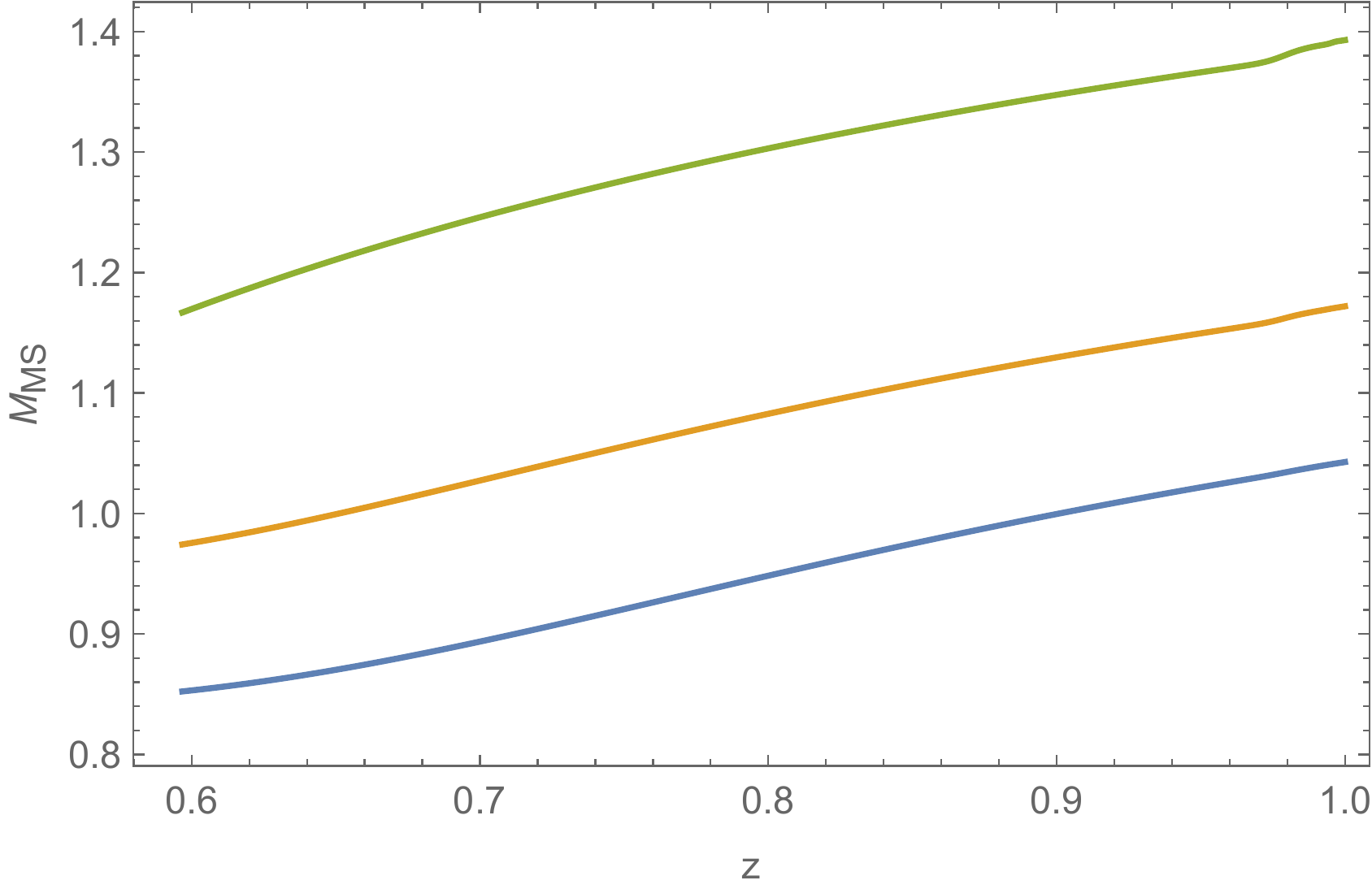}
    \includegraphics[width = 0.4\textwidth]{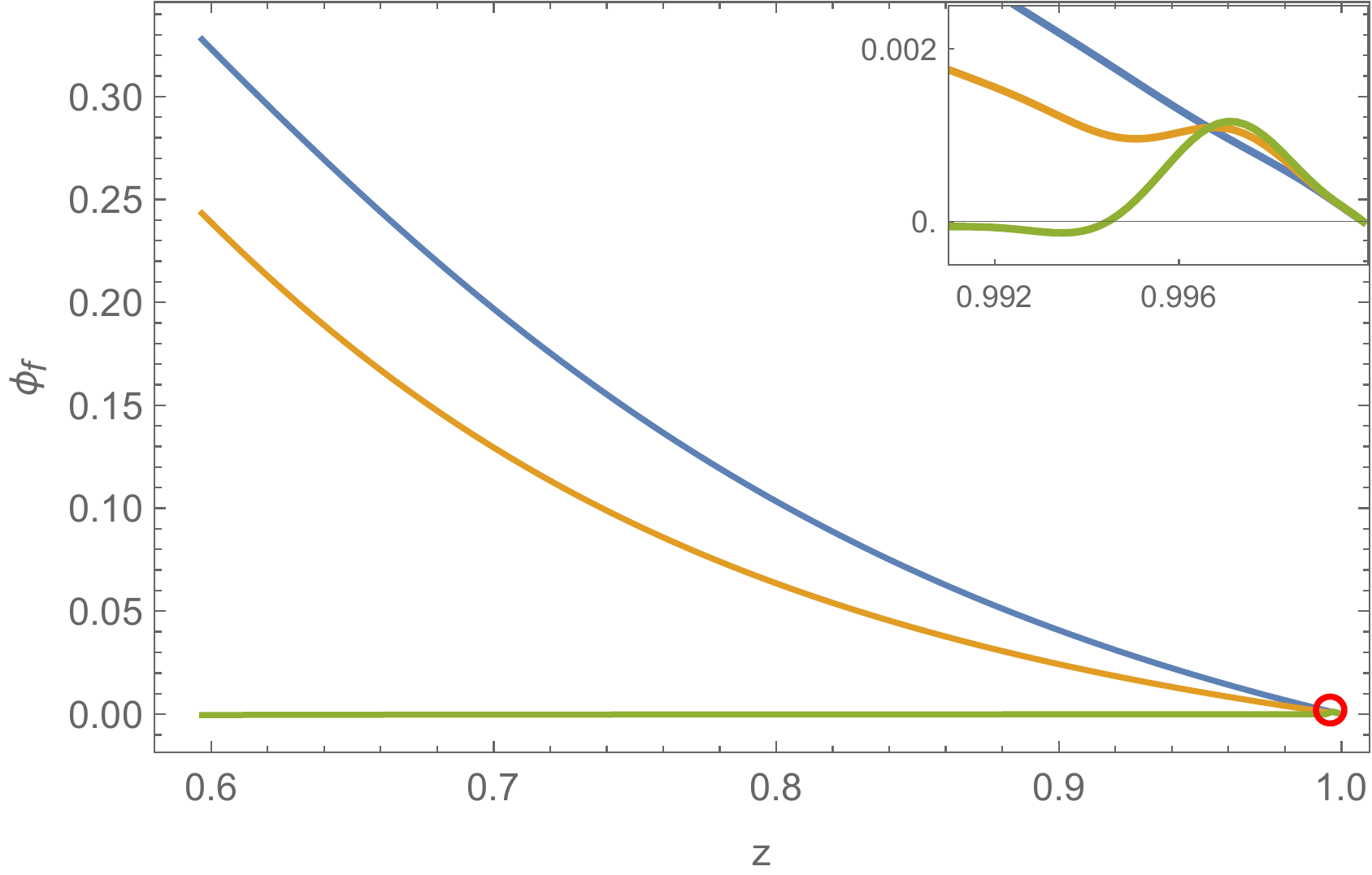}
    \includegraphics[width = 0.4\textwidth]{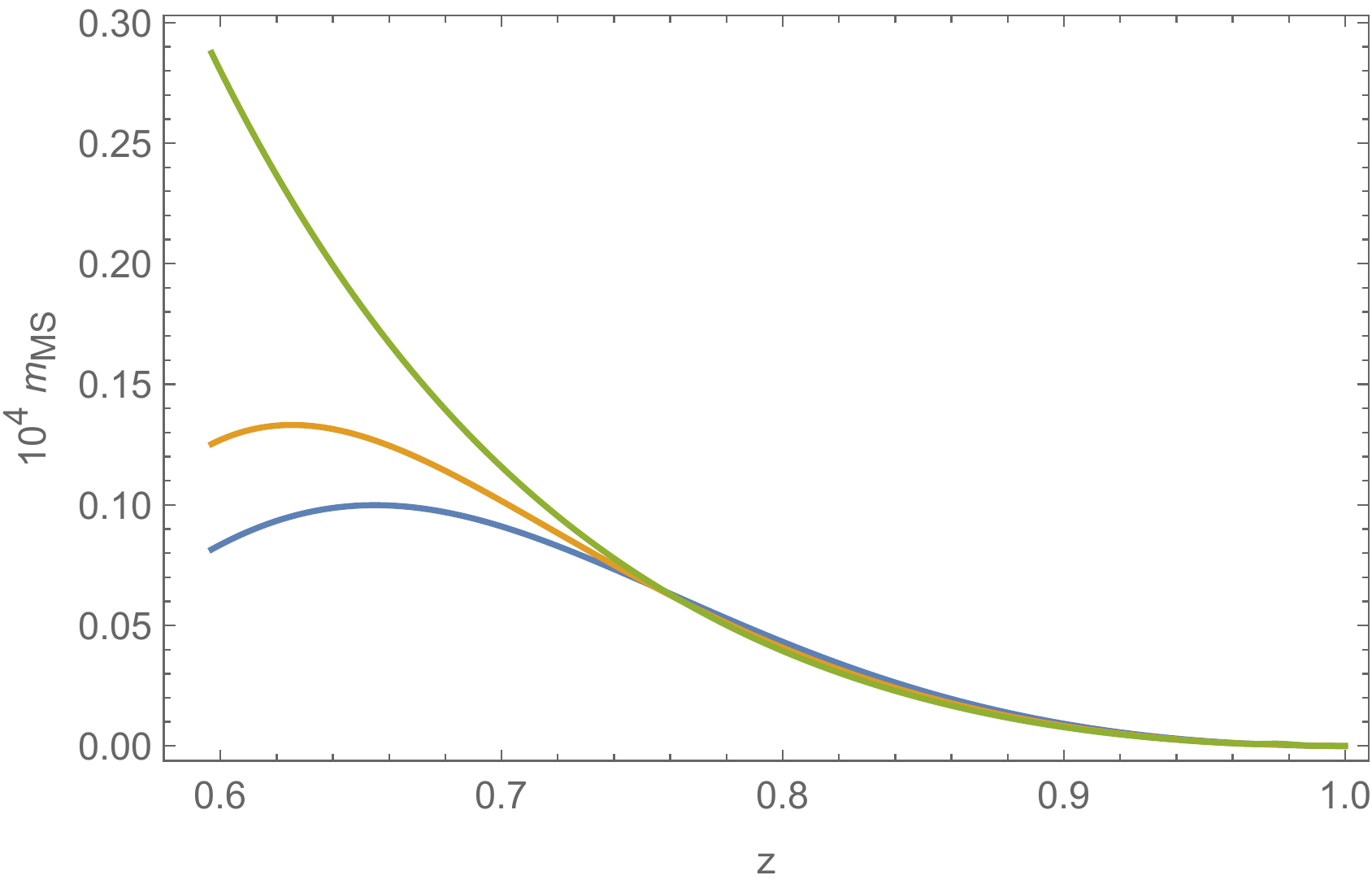}
    \caption{The descalarization for different $B$ but fixed parameters $b=-10,Q=0.8$. The upper left plot shows the evolution of the scalar field at the horizon and the remain plots are the spatial distributions of $M_{\textrm{MS}}$ (upper right), $\phi_{f}$ (bottom left) and $m_{\textrm{MS}}$ (bottom right) respectively at the final state.}
    \label{fig:evolution of descalarization}
\end{figure}

\begin{figure}[htbp]
    \centering
    \includegraphics[width = 0.4\textwidth]{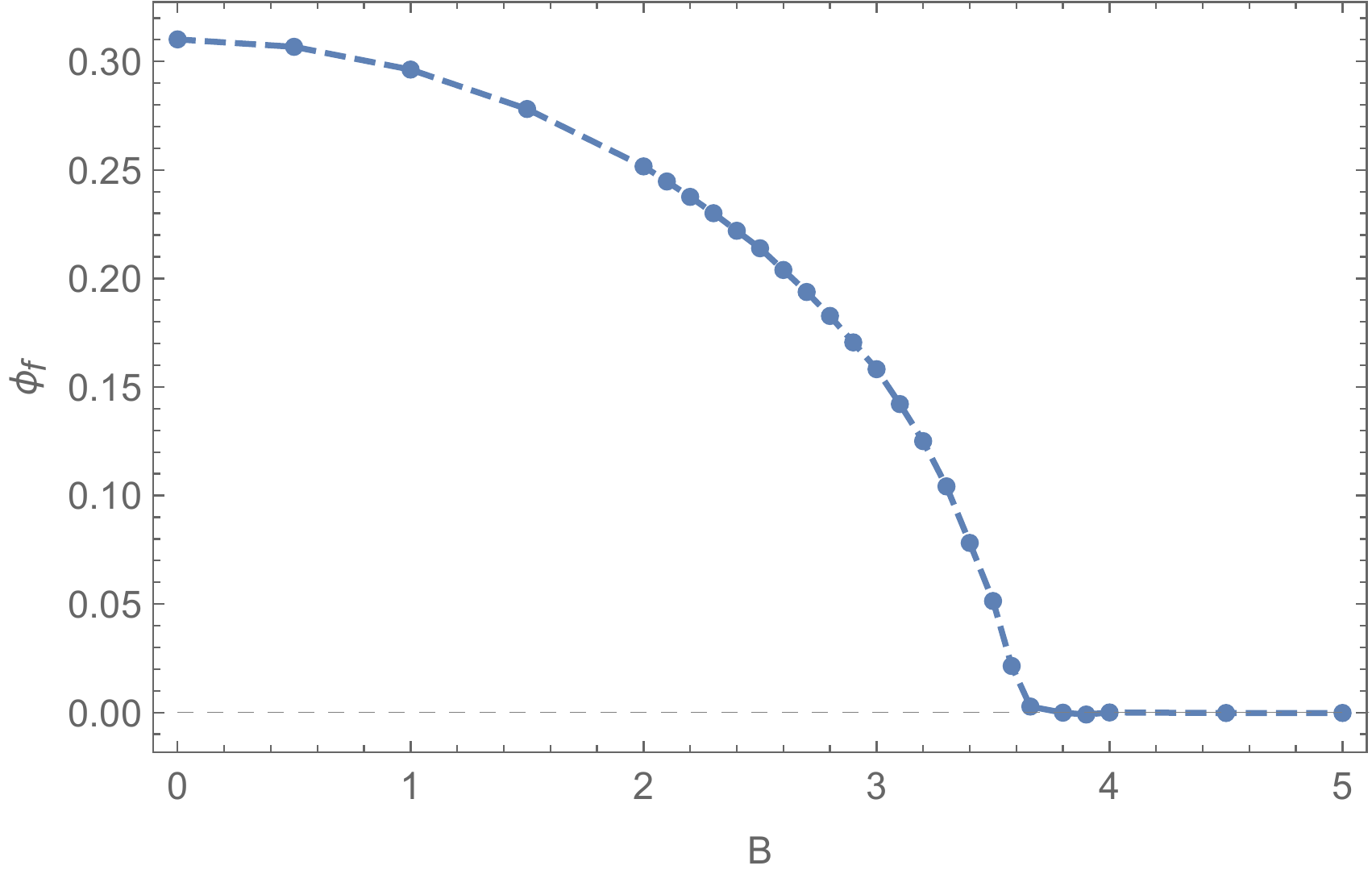}
    \includegraphics[width = 0.4\textwidth]{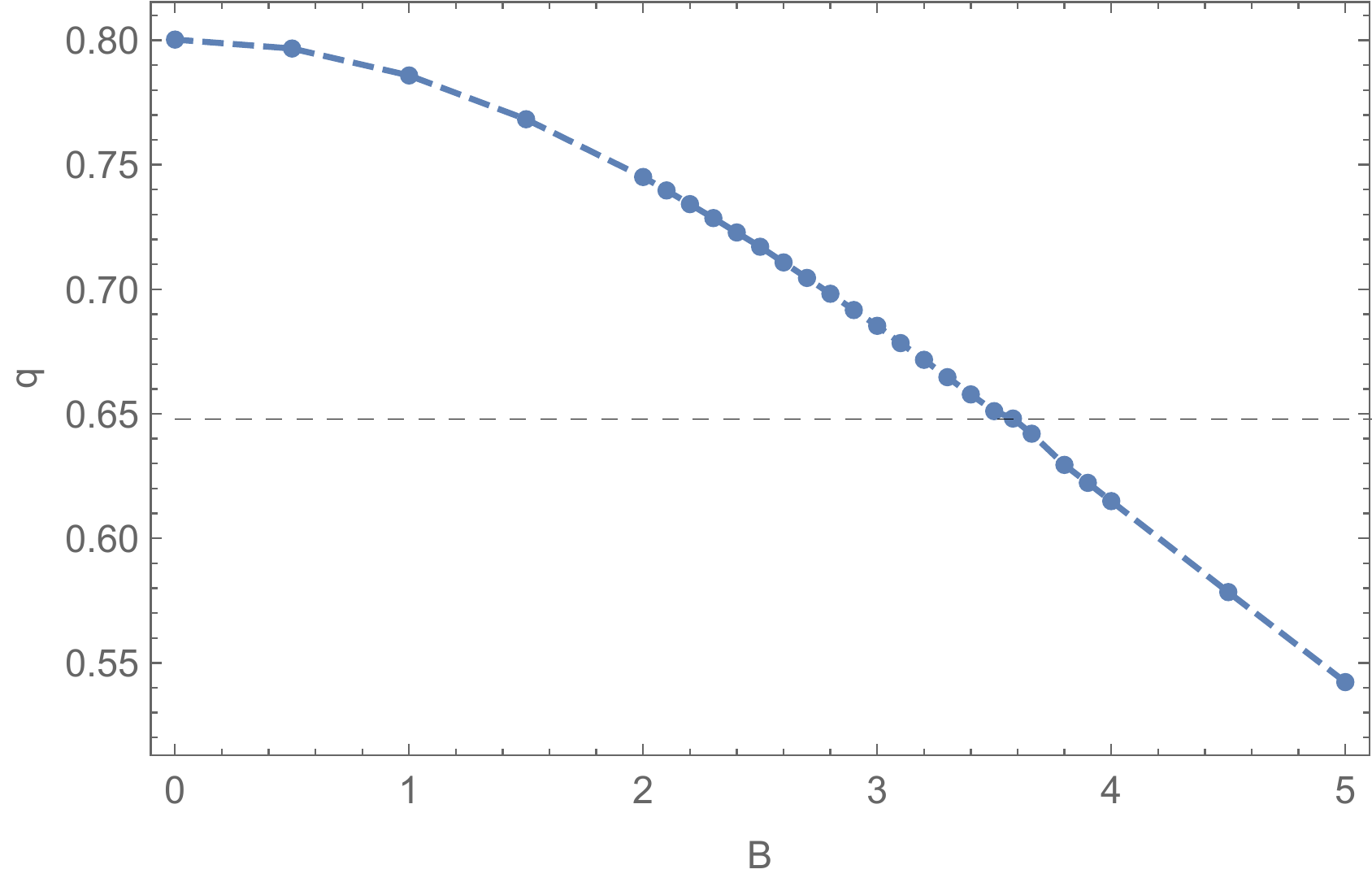}
    \caption{The final value of  scalar field  $\phi_{f}$  on the apparent horizon and effective charge to mass ratio $q=Q/M_s$ versus   $B$ when $b=-10,Q=0.8$. The dashed horizontal line in the right panel corresponds to the threshold for the descalarization. The final black hole becomes bald when $q$ becomes smaller than the threshold value.}
    \label{fig:descalarization vs A}
\end{figure}

Then we investigate the dynamical descalarization under the same set of model parameters ($-b=10$, $Q=0.8$), but with varying strength  $B$. In the upper left panel of Fig.~\ref{fig:evolution of descalarization}, the scalar field decreases with the increase of  $B$ until the scalar hair is removed. This relation is more explicit in the left panel of Fig.~\ref{fig:descalarization vs A}. Note that  the descalarization at the threshold is continuous. This implies that the descalarization is a second order phase transition. The scalar field  distributions of the final state are also displayed in Fig.~\ref{fig:evolution of descalarization}. The red circle in the left bottom panel is enlarged at the upper right corner. The Misner-Sharp mass and its density at late times are also shown. The mass density of the scalarized black hole in the near horizon region is smaller than the descalarized one (the RN black hole). 

In the right plot of Fig.\ref{fig:descalarization vs A} we confirm the conclusions above, the effective parameter $q=Q/M_{\textrm{s}}$, where  $M_{\textrm{s}}$ is the effective mass of the black hole system, determines whether the final state of evolution is scalarized or descalarized. The gray horizontal dashed line in the right panel denotes the threshold value $q_{\textrm{c}}$ for the shedding of scalar hair at $b=-10$. 

In this subsection, we fix the mass of initial black hole $M_0=1$. The total energy of the system increases with the perturbation, as shown in (\ref{eq:MdM}). 
Descalarization is essentially a dynamic process in which the  charge to mass ratio  $q=Q/M_s$ decreases from scalarized region to the scalar-free region, as marked in the Fig.\ref{fig:migration} by two arrows. The crucial point is that the scalarized black hole must absorb extra energy. 
It is noteworthy that we select the effective mass $M_{\textrm{s}}$ instead of the ADM mass $M$ to calculate the charge to mass ratio. The ADM mass, which serves as the total energy of this system at the spatial infinity, can no longer accurately describe  the energy of the black hole system in the dynamical evolution.  Scalar perturbations going outwards to the spatial infinity occupies considerable amount of $M$ and they should be removed  from the total mass $M$.

\subsection{Energy dissipation and scalarization} \label{subsection III.B}

\begin{figure}[htbp]
    \centering
    \includegraphics[width = 0.52\textwidth]{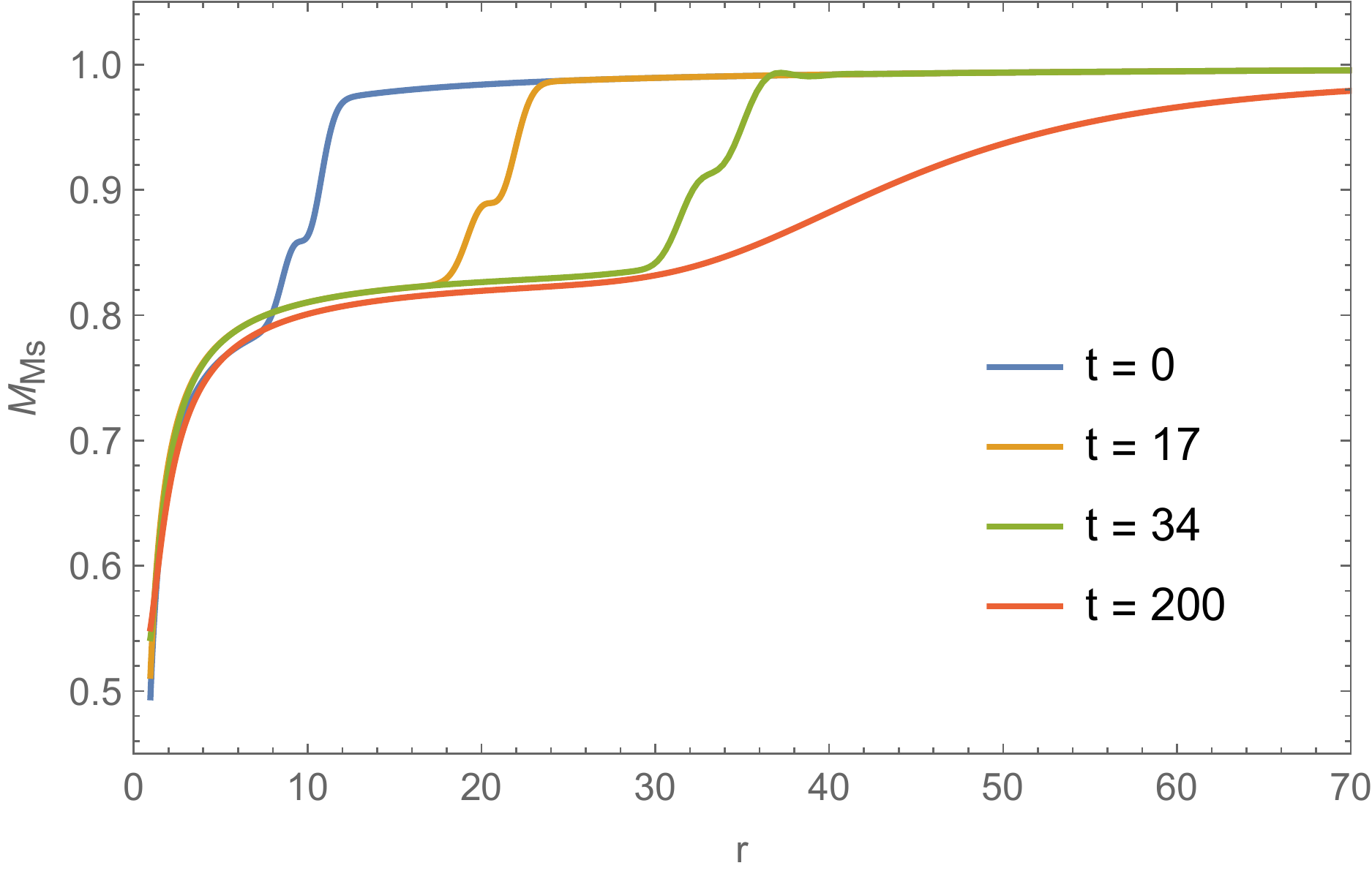}
    \caption{The evolution of the Misner-Sharp mass starting from a RN black hole plus a finite outgoing scalar perturbation when $-b=4,Q=0.8,B=5$. The different color lines represent snapshots of $M_{\textrm{MS}}$ at different time. The total energy $M=1$. But part of the energy escapes to the far region, leaving a black hole system with   charge to mass ratio larger than $Q/M=0.8$, such that the final black hole becomes scalarized.}
    \label{fig:energy dissipation}
\end{figure}

The charge to mass ratio  $q=Q/M_s$  plays a  crucial role in the scalarization/descalarization. In this subsection, we fix the total mass $M=1$ and discuss the effect of energy dissipation to the infinity on the descalarization. The total mass consists of the contribution from the initial RN black hole and the scalar perturbation. 
We fix $b=-4,Q=0.8$ such that a static black hole system with $(-b,Q/M)=(4,0.8)$ lives in the region where no scalarized solution exists, as shown in Fig.\ref{fig:migration}. However, for a dynamical system,  some of the energy inevitably escapes to the infinity such that the effective charge to mass ratio becomes larger than $Q/M$ and the final solution is scalarized.  This is shown in Fig.\ref{fig:energy dissipation}. Here we specify the initial configuration as a scalar-free RN black hole plus a finite outgoing scalar perturbation (\ref{eq:phii}, \ref{eq:Pi}) with $\phi_0=0$. 
  Fig.\ref{fig:energy dissipation} shows that part of the energy of the scalar field transmitted outward to infinity in time evolution. This results in a larger $Q/M_s$ so that the final black hole becomes scalarized. 

\begin{figure}[htbp]
    \centering
    \includegraphics[width = 0.4\textwidth]{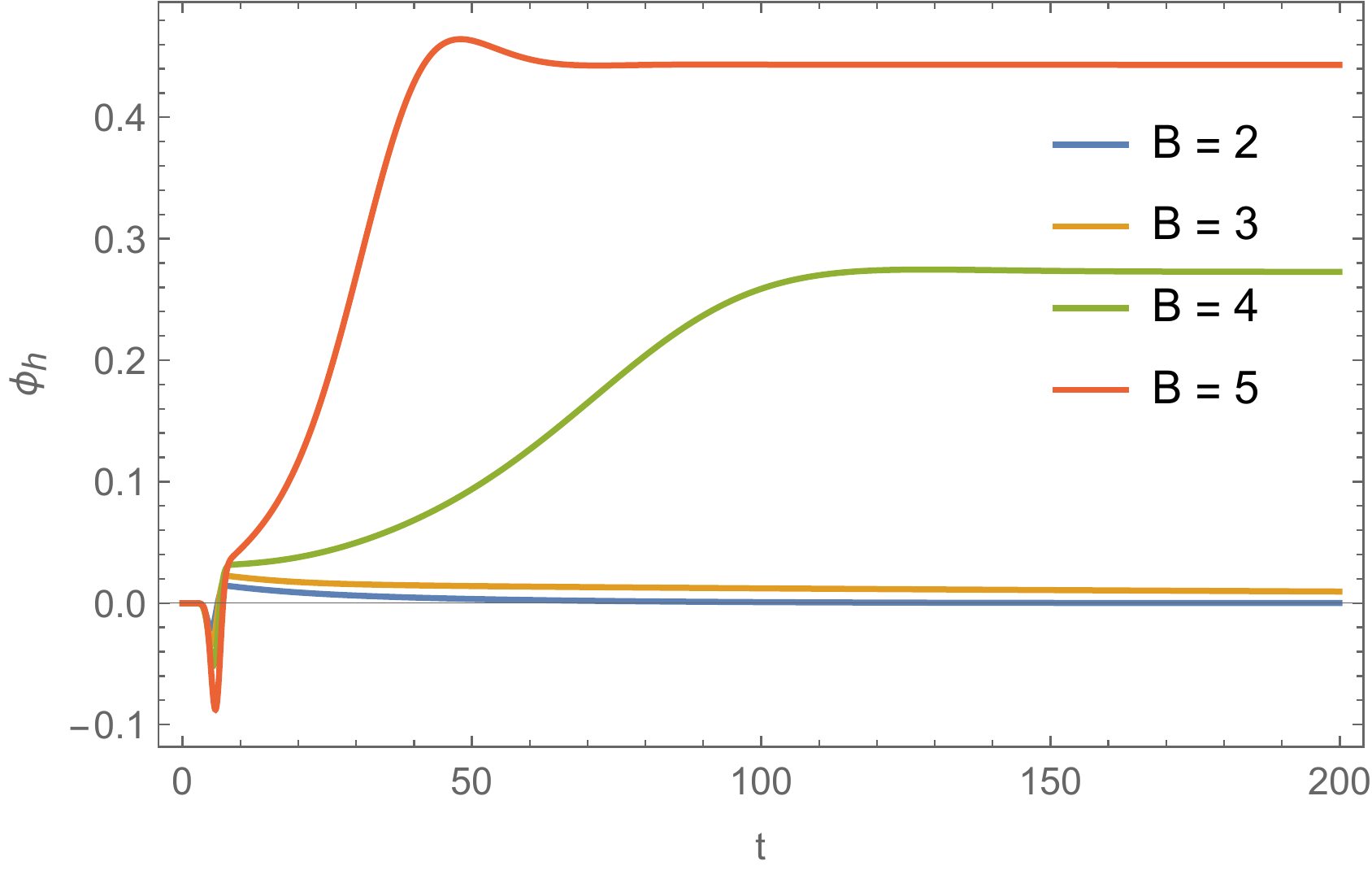}
    \includegraphics[width = 0.4\textwidth]{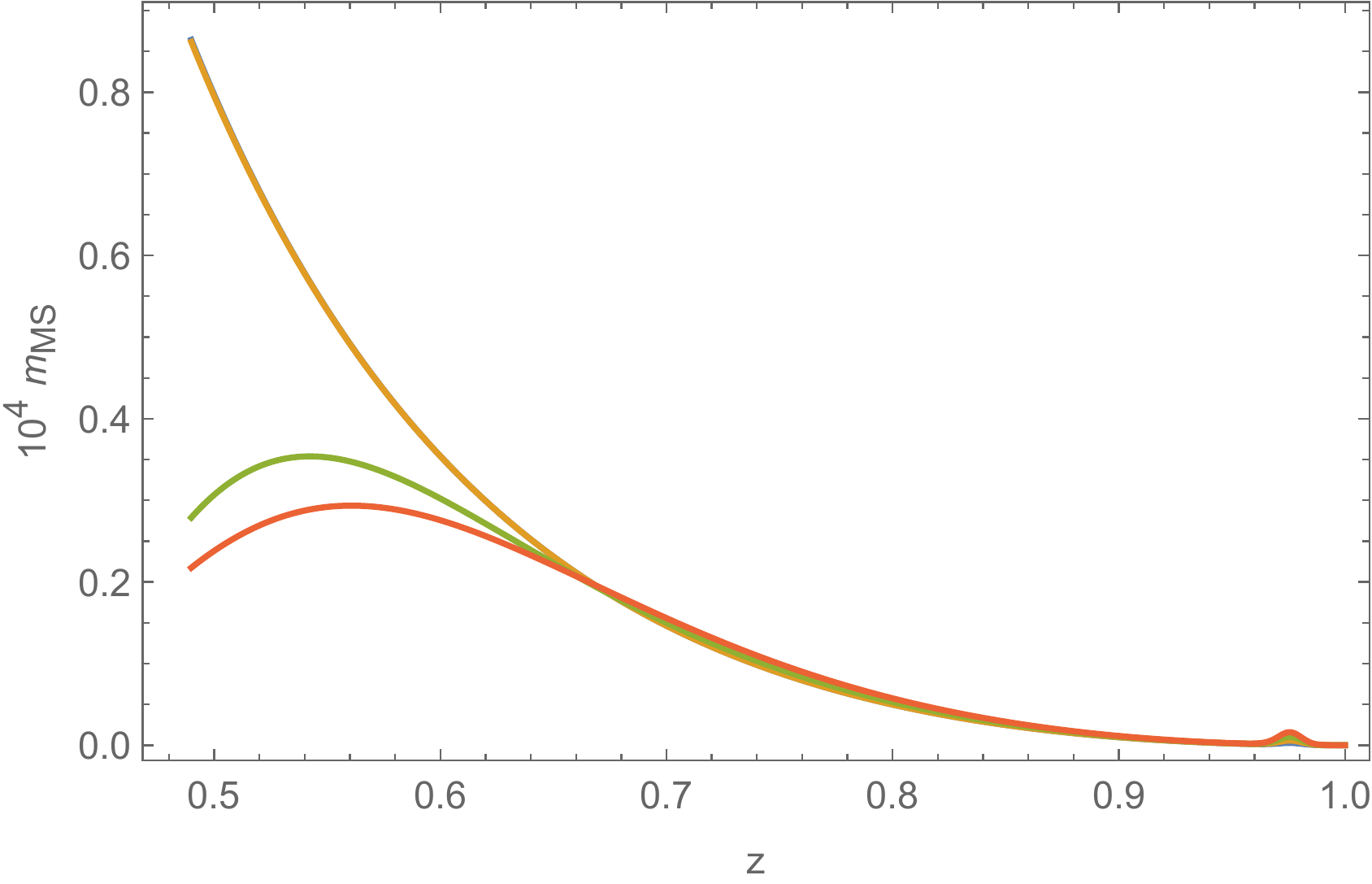}
    \caption{The left plot is the evolution of $\phi_h$ with different $B$ when $-b=4$,$Q=0.8$. The final black hole is scalarized for $B=4,5$. The right plot demonstrates the corresponding mass density $m_{MS}$ at late times. Here the total mass is fixed as $M=1$.}
    \label{fig:scalarization}
\end{figure}

Fig.\ref{fig:scalarization} shows that the scalar hair forms only when $B$, the strength of the initial perturbation,  is   large enough. The difference in the Misner-Sharp mass density between the scalarized black hole and the RN black hole is similar to the previous subsection. We can observe the outward dissipation of  energy of the perturbation in the region near the infinity ($z=1$). It looks small but actually accounts for a considerable proportion of the {total energy because of the compactification $z=\frac{r}{r+1}$ and the long spacing of grid points near the infinity.}  For smaller $B$, the initial RN black hole occupies more energy such that the effective charge to mass ratio is smaller. The final solution can not be scalarized. However, for larger $B$, the initial RN black hole occupies less energy such that the effective charge to mass ratio is larger. The final solution can be scalarized. In other words,  the initial perturbation divides part of the total energy $M=1$. This results in the reduction of the initial energy of the RN black hole, i.e., the effective mass $M_{\textrm{s}}$ of the black hole is less than $M$. With the increase of  $B$, the  $M_{\textrm{s}}$ decreases until $q$ exceeds the critical value $q_{\textrm{c}}$ which eventually leads to the scalarization of black hole. 

\section{Summary and discussion}
\label{section IV}

In this paper, we have investigated the physical process of shedding of scalar hair from scalarized black holes in the EMS theory  in asymptotically flat spacetime. We concentrated on the coupling function $f(\phi)=e^{-b\phi^2}$, which permits the spontaneous scalarization. We started with a scalarized black hole obtained directly from solving the equations of motion. Then we injected scalar perturbations to the spacetime and considered various situations of the injections.  We observed a new physical process in addition to binary black hole merge, which allows a single scalarized black hole to remove its scalar hair after absorbing enough energy.  We find that the effective charge to mass ratio $q$ plays the key role in the descalarization.  Fixing the coupling parameter $b$, the transformation from hairy to bald black holes can only happen when $q$ becomes  small enough.

In Fig.~\ref{fig:descalarization vs A}, we found that the descalarization process is continuous at the threshold so that the phase transition is of the second order. This is consistent with  situation in the AdS spacetime \cite{Zhang:2022cmu}. On the other hand, it was observed that in the EMS theory with higher order coupling function, such as $f(\phi)=e^{-b\phi^4}$, the descalarization can be discontinuous resembling  the first order phase transition in the AdS spacetime \cite{Zhang:2022cmu}. It will be of great interest to study the dynamical descalarization with higher order couplings in asymptotically flat spacetime and examine  the critical behaviors disclosed in  \cite{Zhang2021}. 



\section*{Acknowledgments}
This work is supported by the Natural Science Foundation of China under Grant No. 11805083, 11905083, 12005077, 12075202 and Guangdong Basic and Applied Basic Research Foundation (2021A1515012374).

\end{document}